# Spatial Beam Self-Cleaning and Supercontinuum Generation with Yb-doped Multimode Graded-Index Fiber Taper Based on Accelerating Self-Imaging and Dissipative Landscape

A. NIANG,[1,*] T. MANSURYAN,[2] K. KRUPA,[3] A. TONELLO,[2] M. FABERT,[2] P. LEPROUX,[2] D. MODOTTO,[1] G. MILLOT,[3] V. COUDERC,[2] AND S. WABNITZ[4]

[1] *Dipartimento di Ingegneria dell'Informazione, Università degli Studi di Brescia, via Branze 38, 25123, Brescia, Italy*
[2] *Université de Limoges, XLIM, UMR CNRS 7252, 123 Av. A. Thomas, 87060 Limoges, France*
[3] *Université Bourgogne Franche-Comté, ICB, UMR CNRS 6303, 9 Av. A. Savary, 21078 Dijon, France*
[4] *Dipartimento di Ingegneria dell'Informazione, Elettronica e Telecomunicazioni, Sapienza University of Rome, Via Eudossiana 18, 00184 Rome, Italy*
*[alioune.niang@unibs.it](alioune.niang@unibs.it)*

**Abstract:** We experimentally demonstrate spatial beam self-cleaning and supercontinuum generation in a tapered Ytterbium-doped multimode optical fiber with parabolic core refractive index and doping profile when 1064 nm pulsed beams propagate from wider (120 µm) into smaller (40 µm) diameter. In the passive mode, increasing the input beam peak power above 20 kW leads to a bell-shaped output beam profile. In the active configuration, gain from the pump laser diode permits to combine beam self-cleaning with supercontinuum generation between 520-2600 nm. By taper cut-back, we observed that the dissipative landscape i.e., a non-monotonic variation of the average beam power along the MMF leads to modal transitions of self-cleaned beams along the taper length.

## 1. Introduction

Nonlinear beam propagation in multimode optical fibers (MMFs) has been revisited in recent years: many complex spatio-temporal nonlinear properties have been unveiled [1]. Examples include multimode optical solitons [1-3], geometric parametric instability (GPI) [4], ultra-wide supercontinuum (SC) generation [5-8], spatiotemporal mode-locking [9], and Kerr-induced beam self-cleaning (KBSC) [10-16], to name a few. KBSC results from a multimode four-wave mixing process appearing above a certain threshold peak power, which produces a dramatic reshaping of the output transverse beam profile. In its simplest manifestation, KBSC transforms the output speckled beam pattern into a high-quality, quasi-single mode bell-shaped beam, accompanied by a low power background of higher-order modes (HOMs). KBSC may be accompanied by a complex temporal pulse break-up [13]. Brightness, peak power, and polarization degree of the output beam [16] may all be substantially increased by KBSC. It is important to note that KBSC critically depends on the input beam transverse mode content: launching a tilted beam into the MMF may lead, for example, to KBSC into the $LP_{11}$ mode of the fiber [17]. As a result, for different wave fronts at the fiber input, most of the beam energy remains confined in a low-order mode (LOM) along the MMF.

KBSC has been demonstrated in different MMF types: graded index (GRIN) MMFs [10-13,16], in Ytterbium (Yb) doped MMFs [14] and in photonic crystal non-parabolic refractive index MMFs [15]. Spatial self-cleaning in GRIN MMFs is based on their characteristic spatial beam self-imaging effect, which introduces a longitudinal periodic modulation of the core refractive index, thanks to the Kerr effect. This index modulation acts as a dynamic long period grating, that phase-matches four-wave mixing interactions [18], leading to a complex power transfer between modes, or optical turbulence. This process leads to an irreversible depletion of intermediate modes, accompanied by energy flow into both LOMs and HOMs [19]. This process is analogous to the inverse and direct cascade taking place in 2D hydrodynamic turbulence: the theoretically predicted conservation of the average mode number, during the KBSC process, has been recently experimentally confirmed [20].

As first predicted by Longhi [21], the nonlinear (or dynamic) longitudinal grating induced by self-imaging in GRIN MMFs also leads to the generation of a series of spectral sidebands, ranging from the visible to the near-infrared [1,2,4,22]. SC generation in GRIN MMFs has been observed by injecting either femtosecond or sub-nanosecond pulses in the anomalous (1550 nm) [1,21] or normal (1064 nm) [5-8] dispersion regime, respectively. In the anomalous dispersion regime, spectral broadening results from the interplay between spatiotemporal multimode soliton oscillations [2] and dispersive wave (DW) generation [22]. Spectral broadening leading to red-shifted (with respect to the pump beam) SC is induced by stimulated Raman scattering (SRS) and soliton self-frequency shift. SC to the blue side of the pump is seeded instead by either DWs or GPI (in the normal dispersion regime). Therefore, self-imaging has a crucial role to generate visible light SC in GRIN MMFs.

On the other hand, tapered optical fibers have been shown to exhibit numerous unique advantages over fibers with longitudinally invariant core diameter, including high output beam quality, HOM filtering and broad SC generation [23,24]. Moreover, active rare-earth doped tapers are used to suppress nonlinear effects in chirped pulse amplifiers, when injecting a beam from the small core side, since pulse amplification is accompanied by a progressive decrease of the nonlinear coefficient, owing to the core diameter increase [24]. However, when injecting a beam into the large core side of a multimode fiber taper, since the self-imaging period is directly proportional to the core diameter, accelerated self-imaging occurs, in analogy with the Airy-Talbot effect. Indeed, a recent experiment has shown that accelerated self-imaging in passive, GRIN MMF permits to broaden the spectral width of SC generation on the blue side of the pump [25].

In this work, we study accelerated self-imaging induced nonlinear mode interactions in an active, Yb-doped GRIN multimode fiber taper. We demonstrate visible-mid-infrared SC generation in combination with KBSC in a relatively long (~10 m) active taper, when injecting 500 ps pulses at 1064 nm, propagating in the normal dispersion regime, and from the largest to the smallest taper diameter. We achieve, for the first time to our knowledge, KBSC in a tapered Yb-doped MMF. In addition, we show that the presence of gain induced by a pump laser diode permits to demonstrate KBSC combined with SC generation. Finally, we analyze by the cut-back method the longitudinal evolution of KBSC and supercontinuum along the taper length. This reveals that accelerated self-imaging, combined with a dissipative landscape, may lead to new, unexpected transitions among the self-cleaned modes.

## 2. Experimental set-up

The scheme of the experimental setup is presented in Fig.1. We used a Nd:YAG microchip laser (signal) at 1064 nm with Gaussian spatial beam shape, generating 500 ps pulses at the repetition rate of 500 Hz, with up to 130 kW peak power. A polarizing beam-splitter (PBS) and two half-wave plates (HWPs) were used to adjust the input power (HWP1) and polarization state (HWP2) of the signal.

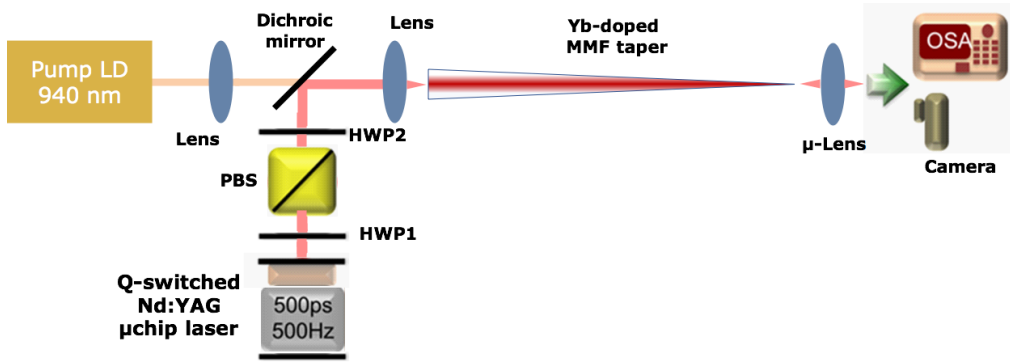

Fig.1. Schematic of the experimental setup for coupling a microchip laser at 1064 nm (signal) and a pump laser diode (LD) into the Yb-doped MMF taper.

In our experiments, we used a 9.5 m long Yb-doped MMF taper exhibiting a strong core absorption at 1064 nm (average attenuation 1.3 dB/m) and parabolic core refractive index (see Fig.2) and doping profiles. The tapered fiber was intentionally wound on a fiber coil which is not presented in Fig1. As shown in Fig.2, the largest input core diameter of the taper was 122 µm (Fig.2(b)) (with a 350x350 µm cladding), whereas the smallest core diameter was close to 37 µm (Fig.2(c)) (with a 90x90 µm cladding). The core diameter was exponentially decreasing along the taper length between these two values (Fig.2(d)). The taper was excited by launching the signal into the largest input diameter. To pump the rare-earth Yb ions, a CW multimode laser diode (LD) of 940 nm and 10 W output power, (Fig.1) was used, providing a net gain for the signal beam propagation along the tapered fiber.

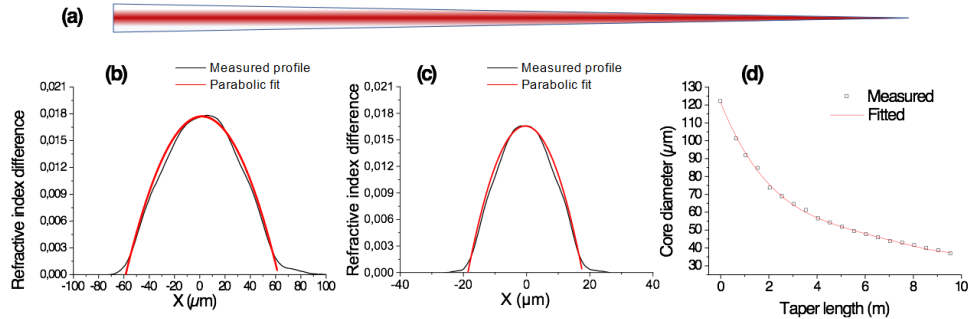

Fig.2. Characteristics of the tapered fiber: (a) doped multimode fiber taper, (b) and (c) profile of the refractive index difference between core and cladding of large and small diameter Yb-doped MMF, respectively, and (d) core diameter of tapered fiber vs taper length.

The taper was placed between two lenses. The first lens with a focal length of 35 mm was placed on a three-axis translation stage, in order to focus both the signal (with a beam diameter of 20 µm at full width of half maximum intensity (FWHMI)) and the pump LD (with a FWHMI beam diameter of 200 µm) into the input face of the MMF. In order to control the input coupling conditions (injection into the fiber), a micro-lens with focal length of 8 mm was used to image the beam from the output face of the MMF (near field) on a CCD camera (Gentec Beamage-CCD12 and Indigo Systems Alpha NIR Camera: 900-1680 nm).We used two Optical Spectrum Analyzers (OSAs) (Ando AQ6315A: 350-1750 nm and Yokogawa AQ6376:1500-3400 nm) to measure spectrum reshaping.

### 3. Experimental results

### 3.1. Pump laser switched off: beam nonlinear self-cleaning

The first experiment was performed in a passive configuration. The signal beam was focused into the core of the Yb-doped GRIN tapered MMFs. The input signal coupled peak power was set to 0.52 kW, and subsequently it was gradually increased up to 114 kW (the damage threshold of the input tapered face). As shown in Fig. 3(a), the spatial beam pattern at the taper output evolved significantly when increasing the signal power: we observed the transition from a speckled beam into a bell-shaped smooth central beam, corresponding to a quasi-single mode emission. The injected beam evolved into the self-cleaned output beam for input peak powers above the 20 kW threshold, and the output beam remained stable for up to 114 kW of coupled signal power. Such behavior can be attributed to KBSC, whereby most of the beam power is transferred into a beam close to the fundamental mode profile ($LP_{01}$) of the fiber [10]. The self-cleaned beam also remained very robust against external disturbances (e.g., intentional bending of the fiber), similarly to the case of passive GRIN MMFs [10], and Yb-MMFs with a non-parabolic refractive index profile [14].

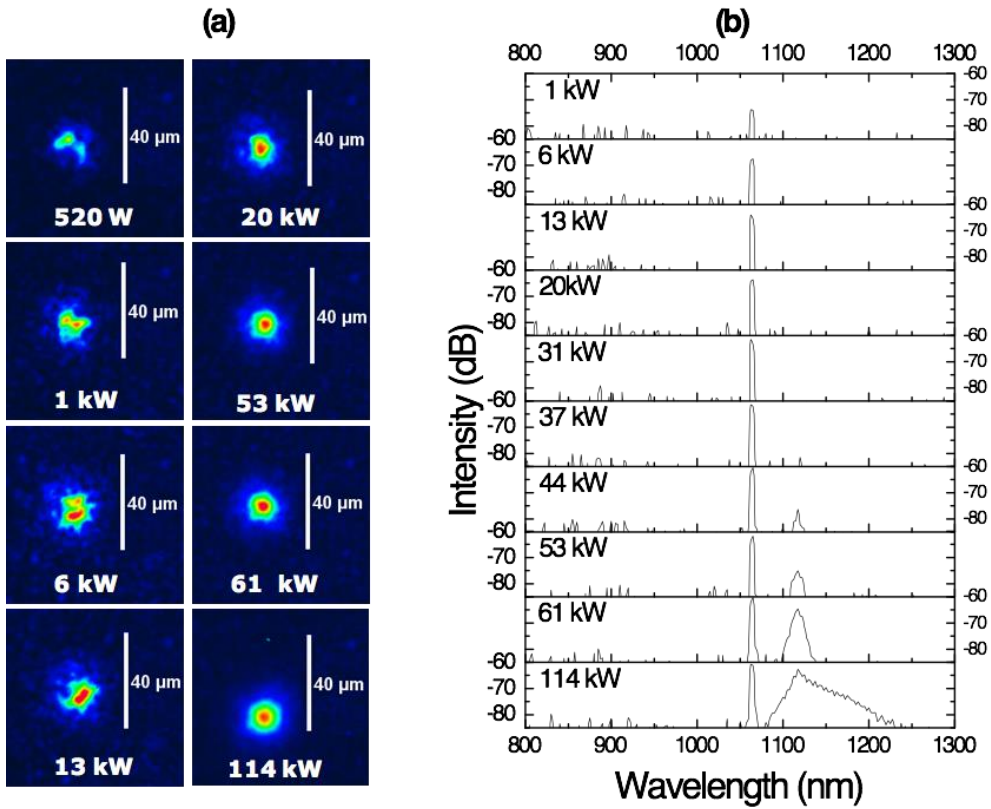

Fig. 3. (a) Near-field spatial distributions and (b) spectra (Ando AQ6315A) of the tapered fiber output beam as a function of input peak power without CW pump. Tapered fiber length was 9.5 m.

Owing to the high residual absorption of the fiber at 1064 nm (total attenuation 12.4 dB), the maximum output peak power was limited to 6.5 kW, for an input power of 114 kW (the damage threshold of the input face of taper). The input power threshold for spatial self-cleaning in our experiment is nearly the same as reported by Guenard et al [14] with a 1.1 m length of lossy (i.e., unpumped) Yb-doped MMF with nearly step index profile and a constant core diameter of 55 µm. Previous experiments of self-cleaning with a passive multimode doped fiber indicated that self-cleaning threshold increases as the fiber length grows larger [14], contrary to the case of lossless GRIN fibers.

Fig.3b shows the output spectra from the tapered MMFs, for different input peak power levels. No significant frequency conversion was observed when progressively increasing the input power, besides the discrete frequency peak appearing above 44 kW, which corresponds to the first Raman Stokes sideband.

### 3.2. Pump laser switched on: beam nonlinear self-cleaning and supercontinuum generation with gain

In a second experiment, we added the CW pump source provided by a 940 nm laser diode delivering up to 10 W, for enabling amplification along the multimode fiber taper. We kept the same 20 μm spot size for the signal laser on the input face of the taper, as in the passive configuration. First, the pump was switched off, and we fixed the signal input peak power at 19.6 kW, just below the KBSC threshold. In this configuration, the transverse content of light at the taper output involved a superposition of the fundamental and higher-order modes, as shown in Fig.4a. Next, we switched on the pump LD, and gradually increased its power, thus adding a growing amount of gain (G) to the fiber. The gain (G) indicated in Fig. 4 corresponds to the ratio between the measured output and input average power of the signal at 1064 nm. Note that the gain G remains low, even for relatively high pump powers. As later discussed with reference to Fig.7, this is due to the dissipative landscape of our nonlinear active taper: the signal is only amplified over the first 2 m of active fiber, and subsequently it is reabsorbed for longer propagation in the taper.

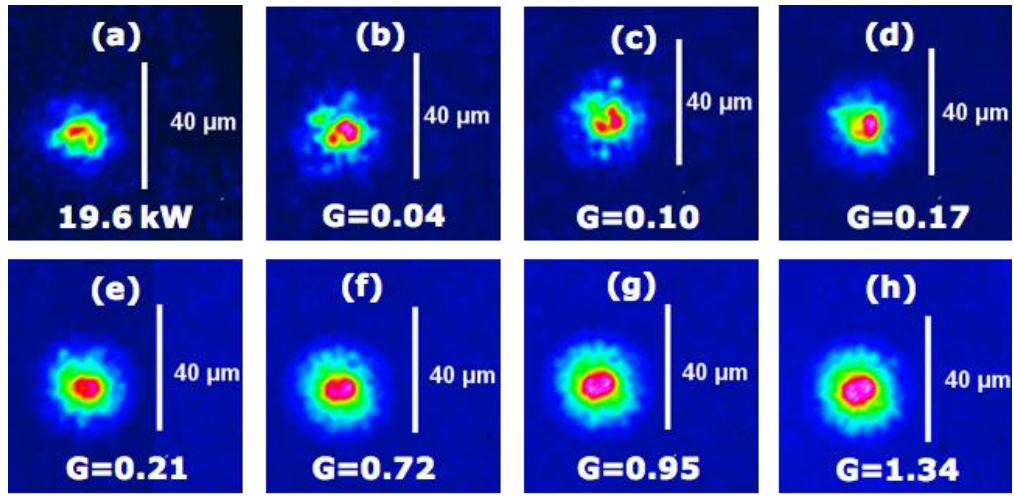

Fig. 4. Near-field spatial distribution of tapered fiber output beams using 19.6 kW input power at 1064 nm. Tapered fiber length was 9.5 m.

In Figs.4b-h we show a series of typical output beam patterns recorded for a fixed input peak power (19.6 kW) of the signal, and different net gain values. We used a bandpass optical filter at 1064 nm with 10 nm bandwidth in front of the camera, in order to block residual radiation from the pump. From Figs.4b-e, a progressive reshaping of the guided beam profile into a bell-shaped cleaned spot can be observed. Such self-cleaned beam started to form for G=0.21, and it remained preserved up to G=1.34, which is the maximum net gain. As discussed before, the limited net gain G is due to pump absorption taking place beyond the first meters of taper, where the pump LD has been fully depleted. Note that our observations clearly show that signal amplification along the taper leads to spatial beam self-cleaning. Besides increasing the effective length, the pump LD leads to a gain guiding mechanism, that cooperates with KBSC for the generation of a bell-shaped output beam profile.

After obtaining KBSC, by further increasing the LD pump power we observed the generation of an ultra-broadband supercontinuum. In order to better understand the evolution of the supercontinuum as a function of gain, we present in Fig.5a typical output spectra for varying gain values. As can be observed, up to G= 0.21 the signal is amplified without showing a significant spectral broadening, besides that induced by the stimulated Raman Stokes peak above G=0.16. By further increasing the gain G, SC generation was obtained starting from G=0.40. From G=0.72, an anti-Stokes sideband induced by GPI is observed, which leads to substantial spectral broadening on the blue side of the signal laser. From Fig.5b, we can observe that for G=1.34 the input signal evolves into a remarkably broad SC spanning between 520 nm and 2600 nm.

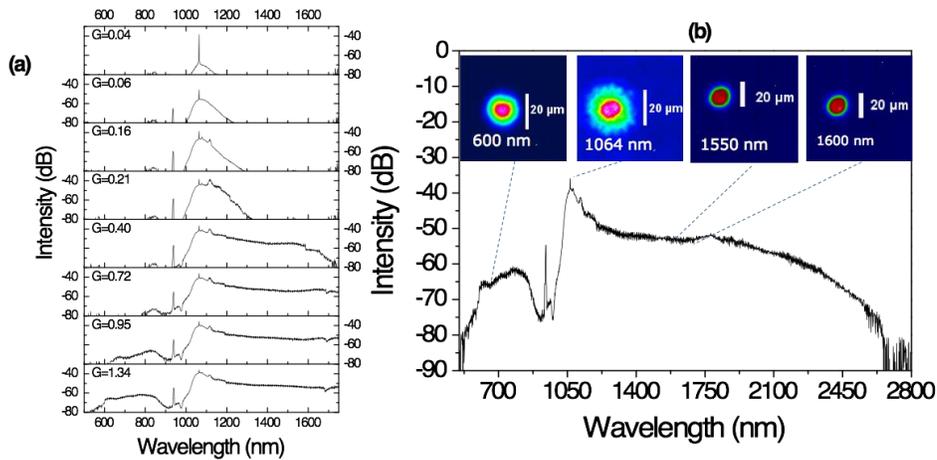

Fig. 5. Spectra obtained from a 9.5 m long taper by using two different OSAs with 19.6 kW input power at 1064 nm. (a) Output spectra (Ando AQ6315A) for different gain values. (b) Supercontinuum spectra (Ando AQ6315A and Yokogawa AQ6376) covering the spectral range 520 nm-2600 nm for maximum gain value of 1.34. The 940 nm peak is a residue from the pump LD. Inset: near-field output beam profile with different bandpass filters.

Subsequently, we characterized the spatial beam profile of the SC from the output face of the taper at various wavelengths, by using bandpass filters with center wavelengths of 600 nm (10 nm bandwidth), 1064 nm (10 nm bandwidth), 1550 nm (12 nm bandwidth) and 1600 nm (12 nm bandwidth), and appropriate imaging cameras. The spatial distributions of the selected parts of the SC are presented in the insets of Fig.5b. At high power levels, the spatial output distributions do not exhibit a speckled structure. Instead, the spatial beam profiles are Gaussian like-shaped at all measured wavelengths across the entire SC spectrum, owing to the interplay of Kerr and Raman self-cleaning, combined with gain guiding. Similar results have been reported on SC generation using passive (i.e., lossless) GRIN multimode fibers [5-8] and tapers [25]. As we shall see, the LD-induced gain introduces a dissipative landscape (i.e., a non-monotonic variation of the average beam power) along the taper, which exacerbates nonlinear effects with respect to the passive taper case, leading to combined SC generation and self-cleaning in the Yb-doped MMF taper.

The SC generation process results from a complex interplay of Raman scattering, soliton-self-frequency shift, dispersive wave generation, and spatiotemporal instabilities (or GPI) of light propagating in GRIN MMFs. On the red-shifted side of the input signal, SRS combined with four-wave mixing is the main mechanism for SC generation. Whereas GPI sidebands provide a seed that fosters subsequent SC generation on the blue side of the signal, between 600 nm and 800 nm, by parametric amplification of the GPI signal. Since the square of the sideband frequency shift approximatively is inversely proportional to the decreasing self-imaging period, accelerating self-imaging in the taper is expected to lead to a progressive blue-shift of GPI sidebands, which largely expands the overall range of spectral broadening. This is obtained at the expense of a decrease in the parametric frequency conversion efficiency, because of the continuous shifting of the phase-matching condition for parametric gain.

In order to analyze the spatial and spectral beam dynamics induced by accelerating self-imaging, we studied the evolution of KBSC and SC generation as a function of taper length. We fixed the input peak power of the signal at 19.6 kW, and set the maximum pumping condition (G=1.34). A cut-back method was used to determine the spatial and spectral beam evolutions along the taper. A 1064 nm optical filter was used at the output of the taper, in order to measure the power of the amplified signal, and to analyze the transverse beam profile. The average power of the amplified signal (at 1064 nm) at the output taper face was measured for different fiber lengths.

Fig.6a summarizes the obtained spectral evolution as a function of taper length. Spectral broadening only appears after the first two meters of fiber, leading to the progressive generation of SC towards the infrared because of the interplay of Raman scattering and Kerr effect. Moreover, a blue shifted continuum seeded by the first anti-Stokes GPI sideband is only clearly visible after 5.55 m of propagation into the taper.

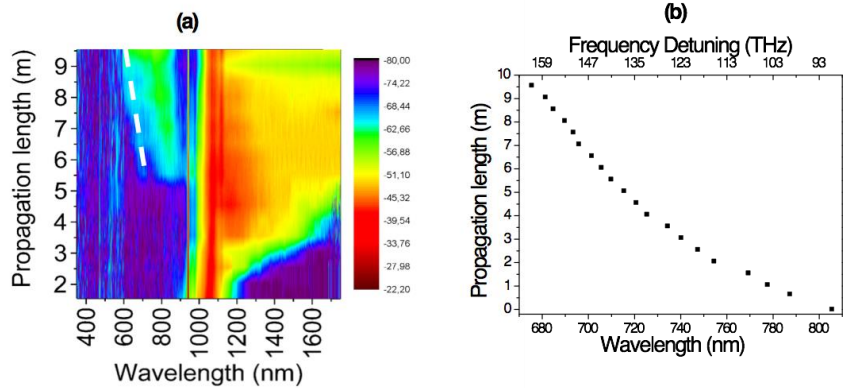

Fig. 6. (a) Experimental spectrum as a function of propagation length in the taper with an input peak power of 19.6 kW at 1064 nm (signal) and maximum pump power (G=1.34). The white dashed line is the analytical prediction of the first anti-Stokes GPI sideband in the tapered fiber between 5.55 m and 9.5 m. (b) Analytical evolution of the first anti-Stokes GPI sideband along the taper.

In order to analytically calculate the frequency position of the sidebands (limited to the first anti-Stokes GPI sideband), we use, according to Ref. [4], the relation $f_h \approx \pm\sqrt{h}f_m$ with h=1,2,3, … and $f_m = (\sqrt{2\pi/(\Lambda\kappa'')})/2\pi$, where $\Lambda$ and $\kappa''$ are the self-imaging period and the group velocity dispersion (GVD), respectively. The self-imaging period varies with the radius $R$ of the fiber core and the relative index difference $\Delta$ as follows: $\Lambda = \pi R/\sqrt{2\Delta}$. Due to the presence of a large number of modes in the tapered fiber, we may consider that the propagation of the amplified signal mainly occurs in bulk silica. Hence, we used the GVD coefficient $\kappa'' = 16.55 \times 10^{-27} s^2/m$ at 1064 nm for a standard GRIN MMFs [4]. Knowing the radius of the core along the taper, we can easily deduce the variation of the self-imaging period along the length of the taper. Therefore, the frequency detuning of the first resonant GPI sideband can be analytically estimated via $f_1$. The frequency detuning and the wavelength of the first anti-Stokes GPI sideband are shown in Fig.6b. From Fig.6b, we can see that the sideband frequency shift increases (and its wavelength decreases) along the taper.

Fig.6a shows that in our experiments, the GPI-generated spectrum only appears between 5.55 and 9.5 m. In this fiber length range, the first GPI sideband shift varies from $f_1 = 137\ THz$ at a core taper radius of ~25 µm (taken at the distance of 5.55 m) to $f_1 = 159\ THz$ at a core taper radius of ~18.5 µm (taken at the distance of 9.5 m). The corresponding wavelength of the first anti-Stokes GPI sideband decreases from 710 nm to 675 nm, as shown by the white dashed curve superimposed with the experimental spectrum in Fig 6.a: the estimated spectral shift at distances between 6 and 9 m is larger than the one observed experimentally. This indicates that the blue-shifting SC is mainly seeded from quantum noise by accelerating self-imaging induced GPI, which occurs over the first 2-3 m of taper, that is before a substantial spectral broadening of the signal occurs. In fact, Fig.6b shows that the predicted GPI peak gain over the first 3 m of taper sweeps across the observed range of blue SC, namely, the 650-800 nm spectral region.

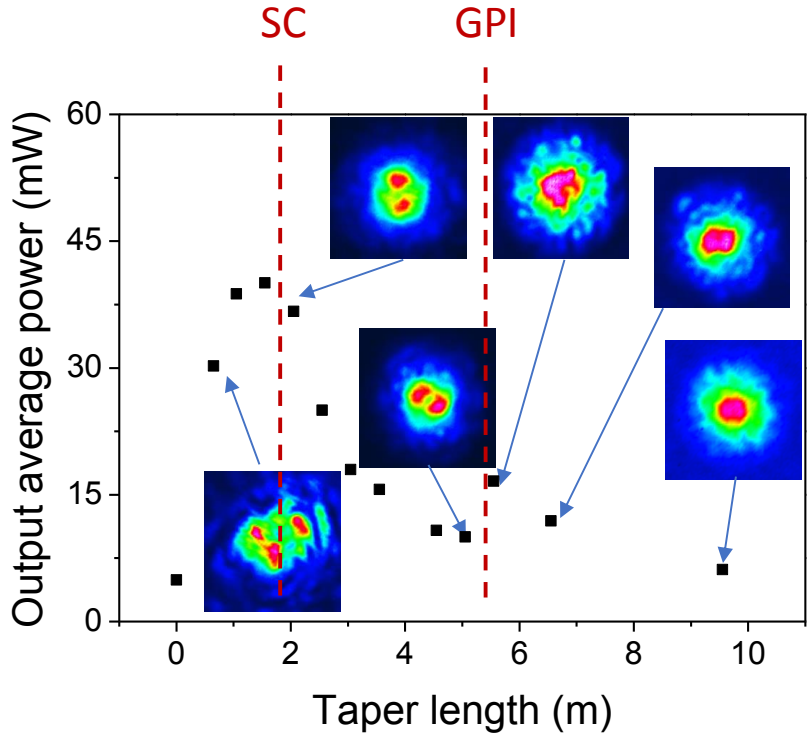

Fig.7. Output average power at 1064 nm as function of taper length showing beam self-cleaning evolution with an input peak power of 19.6 kW (signal) and maximum pump power (G=1.34). Insets: output near-field spatial distributions along the taper.

In order to reveal the dissipative landscape of the taper, we measured output average power of the amplified signal at 1064 nm as a function of taper length, as illustrated in Fig.7. As can be seen, the optimal taper length, *i.e.*, the length at which the amplification of the input signal is maximum corresponds to 1.55 m. This leads to an enhancement of nonlinear effects after 2 m of taper, which effectively turns on Raman gain and the associated spectral broadening.

The nonlinear dynamics of the spatial profile of the beam along the taper induced by the dissipative landscape is shown by the various insets of Fig.7. The spatial beam distribution is speckled at the beginning (first meter) of the propagation, but very interestingly, Fig.7 unveils that the beam is progressively self-cleaned into different LOMs during its propagation along the taper. Between 2 m and 6 m, that is in the region where SC is generated, but before the appearance of a GPI-induced spectrum, a LOM (which resembles the $LP_{11}$ mode) is generated. Subsequently, for distances above 6 m, and in correspondence with the appearance and broadening of the GPI spectrum, self-cleaning occurs into a bell-shaped beam, whose size is close to that of the fundamental $LP_{01}$ mode. Therefore, in the presence of a dissipative landscape (i.e., the interplay of gain and loss along the MMF), self-cleaning into an $LP_{11}$-like mode appears as a transient effect. The thresholds of GPI and SC generation are inserted in Fig7 as vertical dashed red lines at the corresponding fiber positions.

### 4. Conclusions

To conclude, we experimentally demonstrated, for the first time to our knowledge, that tapered active ytterbium-doped multimode fibers with a parabolic index and doping profile may provide a new and versatile platform for high beam quality supercontinuum generation ranging from the visible to the mid-infrared, when pumped in the normal dispersion regime at 1064 nm. The interplay of GPI and SRS allowed us to generate, in combination with a gain/loss landscape, a spectral bandwidth extending from 520 nm up to 2600 nm by using a 9.5 m long tapered fiber. Accelerating self-imaging led to Kerr-self beam cleaning in both passive and active configurations for our tapered Yb-doped GRIN MMF. In the active case, cooperation of KBSC and Raman beam cleanup led to high beam quality emission across the entire SC bandwidth.

By the cut-back method we studied the evolution of beam self-cleaning and supercontinuum generation along the tapered fiber operating in the active configuration. We observed that the output spatial distribution of the beam evolves from speckles in the first meters, into a dual lobe, $LP_{11}$-like mode as SC generation is obtained, and finally into a bell-shaped beam close to the fundamental mode as the GPI-induced spectrum is developed.

Active MMF tapers may thus combine accelerating self-imaging with a dissipative landscape, and permit a versatile control of the spectral and spatial content of multimode light beams. These results may find important applications in multimode fiber lasers and in nonlinear imaging technologies.

### Funding

We acknowledge financial support of: iXcore research foundation; French program 364 "Investissement d'Avenir" Project No. ISITE-BFC-299365 (ANR-15 IDEX-0003); the European Research Council (ERC) under the European Union's Horizon 2020 research and innovation program (grant No. 740355); CILAS company (Ariangroup).

### Acknowledgments

We would like to thank the Fiber Optics Research Institute (FORC) in Moscow for fabricating the tapered Yb-doped MMF.

### References

1. L. G. Wright, D. N. Christodoulides, and F. W. Wise, "Controllable spatiotemporal nonlinear effects in multimode fibres," Nat. Photon. **9**, 306 (2015).
2. W. H. Renninger and F. W. Wise, "Optical solitons in graded-index multimode fibres," Nat. Commun. **4**, 1719 (2013).
3. L. G. Wright, W. H. Renninger, D. N. Christodoulides, and F. W. Wise, "Spatiotemporal dynamics of multimode optical solitons," Opt. Express **23**, 3492–3506 (2015).
4. K. Krupa, A. Tonello, A. Barthélémy, V. Couderc, B. M. Shalaby, A. Bendahmane, G. Millot, and S. Wabnitz, "Observation of geometric parametric instability induced by the periodic spatial self-imaging of multimode waves," Phys. Rev. Lett. **116**, 183901 (2016).
5. G. Lopez-Galmiche, Z. Sanjabi Eznaveh, M. A. Eftekhar, J. Antonio Lopez, L. G. Wright, F. Wise, D. Christodoulides, and R. Amezcua Correa, "Visible supercontinuum generation in a graded index multimode fiber pumped at 1064nm," Opt. Lett. **41**, 2553–2556 (2016).
6. K. Krupa, C. Louot, V. Couderc, M. Fabert, R. Guenard, B. M. Shalaby, A. Tonello, A. Barthélémy, D. Pagnoux, P. Leproux, A. Bendahmane, R. Dupiol, G. Millot, and S. Wabnitz, "Spatiotemporal characterization of supercontinuum extending from the visible to the mid-infrared in multimode graded-index optical fiber," Opt. Lett. **41**, 5785-5788 (2016).
7. M. A. Eftekhar, L. G. Wright, M. S. Mills, M. Kolesik, R. Amezcua Correa, F. W. Wise, and D. N. Christodoulides, "Versatile supercontinuum generation in parabolic multimode optical fibers," Opt. Express **25**, 9078-9087 (2017).
8. U. Teğin, and B. Ortaç, "Cascaded Raman scattering based high power octave-spanning supercontinuum generation in graded-index multimode fibers," Sci. Rep. **8,** 30252-9 (2018).
9. L. G. Wright, D. N. Christodoulides, and F. W. Wise, "Spatiotemporal mode-locking in multimode fiber lasers," Science **358**, 94-97 (2017).
10. K. Krupa, A Tonello, B. Shalaby, M. Fabert, A. Barthélémy, G. Millot, S. Wabnitz, and V. Couderc, "Spatial beam self-cleaning in multimode fiber," Nat. Photon. **11**, 237-241, (2017).
11. Z. Liu, L. G. Wright, D. N. Christodoulides, and F. W. Wise, "Kerr self-cleaning of femtosecond-pulsed beams in graded-index multimode fiber," Opt. Lett. **41**, 3675-3678 (2016).
12. L. G. Wright, Z. Liu, D. A. Nolan, M.-J. Li, D. N. Christodoulides, and F. W. Wise, "Self-organized instability in graded index multimode fibres," Nat. Photon. **10**, 771–776 (2016).
13. K. Krupa, A. Tonello, V. Couderc, A. Barthélémy, G. Millot, D. Modotto, and S. Wabnitz, "Spatiotemporal light-beam compression from nonlinear mode coupling," Phys. Rev. A **97**, 043836 (2018).
14. R. Guenard, K. Krupa, R. Dupiol, M. Fabert, A. Bendahmane, V. Kermene, A. Desfarges-Barthelemot, J. L. Auguste, A. Tonello, A. Barthélémy, G. Millot, S. Wabnitz, andV. Couderc, "Kerr self-cleaning of pulsed beam in ytterbium doped multimode fiber," Opt. Express **25**, 4783-4792 (2017).
15. R. Dupiol, K. Krupa, A. Tonello, M. Fabert, D. Modotto, S. Wabnitz, G. Millot, S. Wabnitz, and V. Couderc, "Interplay of Kerr and Raman beam cleaning with a multimode microstructure fiber," Opt. Lett. **43**, 587-590 (2018).
16. Katarzyna Krupa, Graciela Garmendia Castañeda, Alessandro Tonello, Alioune Niang, Denis S. Kharenko, Marc Fabert, Vincent Couderc, Guy Millot, Umberto Minoni, Daniele Modotto, and Stefan Wabnitz, "Nonlinear polarization dynamics of Kerr beam self-cleaning in a GRIN multimode optical fiber," Opt. Lett. **44**, 171-174 (2019).
17. E. Deliancourt, M. Fabert, A. Tonello, K. Krupa, A. Desfarges-Berthelemot, V. Kermene, G. Millot, A. Barthélémy, S. Wabnitz, and V. Couderc, "Kerr beam self-cleaning on the $LP_{11}$ mode in graded-index multimode fibers," OSA Continuum **2**, 1089-1096 (2019).
18. Martin Schnack, Tim Hellwig, Maximilian Brinkmann, and Carsten Fallnich, "Ultrafast two-color all-optical transverse mode conversion in a graded-index fiber," Opt. Lett. **40**, 4675-4678 (2015).
19. P. Ascheri, G. Garnier, C. Michel, V. Doya and A. Picozzi. "Condensation and thermalization of classical optical waves in a waveguide" Phys. Rev. A, 83, 033838 1-13 (2011).
20. E. V. Podivilov, D. S. Kharenko, V. A. Gonta, K. Krupa, O. S. Sidelnikov, S. Turitsyn, M. P. Fedoruk, S. A. Babin, and S. Wabnitz, "Hydrodynamic 2D turbulence and spatial beam condensation in multimode optical fibers," Phys. Rev. Lett. **122**, 103902 (2019).
21. S. Longhi, "Modulational instability and space–time dynamics in nonlinear parabolic-index optical fibers," Opt. Lett. **28**, 2363-5 (2003).
22. L. G. Wright, S. Wabnitz, D. N. Christodoulides, and F. W. Wise, "Ultrabroadband dispersive radiation by spatiotemporal oscillation of multimode waves," Phys. Rev. Lett. **115**, 223902 (2015).
23. T. Birks, W. J. Wadsworth, and P.S.J. Russell, "Supercontinuum generation in tapered fibers," Opt. Lett. **25**, 1415 (2000).
24. C. Shi, X. Wang, P. Zhou, X. Xu, and Q. Lu, "Theoretical study of mode evolution in active long tapered multimode fiber," Opt. Express **24**, 19473-19490 (2016).
25. Y. Lumer, L. Drori, Y. Hazan, and M. Segev, "Accelerating self-imaging: the Airy-Talbot effect," Phys. Rev. Lett. **115**, 013901 (2015)
26. M. A. Eftekhar, Z. Sanjabi-Eznaveh, J. E. Antonio-Lopez, H. Aviles, S. Benis, M. Kolesik, A. Schülzgen, F. W. Wise, R. Correa, and D. N. Christodoulides, "Accelerating nonlinear interactions in tapered multimode fibers," in Conference on Lasers and Electro-Optics, Optical Society of America, paper FTh1.M3 (2018).